**Accepted Version**

Publication date: 14th April 2023

Embargo: No Embargo (Accepted Version, under conditions), 24 Months (Published Version)

European Union, Horizon 2020, Grant Agreement number: 857470 — NOMATEN — H2020-WIDESPREAD-2018-2020

DOI: https://doi.org/10.1016/j.saa.2023.122341

### Spectroscopic studies on phosphate-modified silicon oxycarbide-based amorphous materials

*Magdalena Gawęda[1]\*, Piotr Jeleń[2], Maciej Bik[2], Magdalena Szumera[2], Zbigniew Olejniczak[3], Maciej Sitarz[2]*

[1] *NOMATEN CoE, NOMATEN MAB, National Centre for Nuclear Research, A. Soltana 7 Str., 05-400 Otwock-Świerk, Poland,*

[2] *AGH University of Science and Technology, Faculty of Materials Science and Ceramics, A. Mickiewicza 30 Av, 30-059 Kraków, Poland*

[3] *The Henryk Niewodniczański Institute of Nuclear Physics, Polish Academy of Sciences, Radzikowskiego 152 Str., 31-342 Kraków, Poland*

*\*corresponding author, email: Magdalena.Gaweda@ncbj.gov.pl*

**Abstract**

Vibrational spectroscopy is the most effective, efficient and informative method of structural analysis of amorphous materials with silica matrix and, therefore, an indispensable tool for examining silicon oxycarbide-based amorphous materials (SiOC). The subject of this work is a description of the modification process of SiOC glasses with phosphate ions based on the structural examination including mainly Infrared and Raman Spectroscopy. They were obtained as polymer-derived ceramics based on ladder-like silsesquioxanes synthesised via the sol-gel method. With the high phosphate's volatility, it was decided to introduce the co-doping ions to create [AlPO$_4$] and [BPO$_4$] stable structural units. As a result, several samples from the Si*P*OC, Si*PAl*OC and Si*PB*OC systems were obtained with various quantities of the modifiers. All samples underwent a detailed structural evaluation of both polymer precursors and ceramics after high-temperature treatment with Fourier-transformed infrared spectroscopy (FTIR), Raman spectroscopy, X-ray diffraction (XRD) and magic angle spinning nuclear magnetic resonance (MAS-NMR). Obtained results proved the efficient preparation of desired materials that exhibit structural parameters similar to the unmodified one. They were X-ray-amorphous with no phase separation and crystallisation. Spectroscopic measurements confirmed the presence of the crucial Si-C bond and how modifying ions are incorporated into the SiOC network. It was also possible to characterise the turbostratic free carbon phase. The modification was aimed to improve the bioperformance of the materials in the context of their future application as bioactive coatings on metallic implants.

### 1. Introduction

Silicon oxycarbide (usually abbreviated to SiOC) is a material of neither fully defined nor unified structure. As the literature data shows, it is due to the strong dependence of structure on the preparation method and eventual ionic modifications. The unique structure is a superposition of amorphous silica (v-SiO$_2$) and silicon carbide (SiC) with an accompanying turbostratic carbon phase [1]. Two well-established composite models argue what constitutes the SiOC matrix and reinforcement – amorphous silica or graphite-like carbon [2, 3]. Nevertheless, the common factor is the chemical bonds between the two phases, more precisely, between silicon and carbon. The Si-C bond's presence is responsible for unique properties of SiOC-based materials, which are the combination of properties of amorphous silica and silicon carbide. They possess good mechanical properties, thermal and chemical stability, corrosion and oxidation resistance, good adhesion to metallic substrates, biocompatibility and bioactivity [4, 5, 6]. It has been proven as a very versatile material since it might be applied as a bulk matrix for Li-ion batteries [7] or protective coating against high-temperature conditions [8, 9, 10], humidity [11], and saltwater solutions [12]. Despite already proven parameters sufficient for implantology, further material development is necessary for better or more targeted properties [5, 13].

The premise of this work stems from the lack of knowledge on silicon oxycarbide modification with phosphorus ions. However, modifications with other bioelements, such as calcium and magnesium, are on a high level of development described in the literature [14]. Phosphate ions are one of the most often used ions to modify classic bioglasses. This work is highly focused on the possibility of modification of SiOC with phosphate ions despite the well-known volatile character of its compounds [15]. Hence the three aspects of the material's preparation were considered: (1) the application of different derivatives, especially of higher molecular weight, (2) the alteration of the synthesis parameters and (3) the use of other ions securing bonding phosphate ions in a silica-based network. Since the first two aspects are relatively straightforward, the last one needs more explanation. Implementation and maintaining phosphorus is a well-known critical problem also in classic melted glass. As the solution, implementation of other ions was applied: aluminium and boron. Both of them have the ability to create [AlPO$_4$] and [BPO$_4$] structural units with dimensions equivalent to the [SiO$_4$] units [16, 17]. Since they can easily build into the silica network and the network of SiOC amorphous materials is based on amorphous silica, the same approach was applied in work. However, additional ions present in the material will impact its structure and properties. As an example, the presence of boron ions might induce significant changes in the free carbon phase by creation of preferential [BC$_x$O$_{3-x}$] units [18]. Since this article focuses on the structural investigation, functional properties (such as quality of the coating, corrosion resistance, bioactivity and biocompatibility) of SiOC-based materials modified with phosphate ions will be the topic of further study.

## 2. Materials and methods

Ladder-like silsesquioxanes modified with phosphate, alumina and boron ions were synthesised via the sol-gel method. Acidic hydrolysis and polycondensation were conducted based on the methodology previously developed for pure (nonmodified) SiOC preceramic precursors [4, 5] as well as previous experiences with modifications with alumina [12, 19] and cerium ions [20]. The low reactivity of phosphate ions [15] determined the pre-trial of acidic hydrolysis and condensation of the chosen compounds containing phosphorus: (2-diethylphosphatoethyl)-triethoxysilane and triethyl phosphite. The first of the mentioned compounds already contained a silicon atom in its molecule and was chosen as the presumably more effective alternative for the simpler compound (second mentioned). Such prepared compounds were added to the solution containing alkoxysilanes: methyltriethoxysilane (with T structural unit: [SiO$_3$C]) and dimethyldiethoxysilane (D units: [SiO$_2$C$_2$]) in the 2:1 molar ratio. In the case of the co-modified precursors, before the addition of the pre-condensed phosphate compound, the aluminium or boron-containing compounds were added. The addition of alumina was challenging due to the opposite reason - very high reactivity. It was introduced to the solution in the chelated form obtained with acetyl acetate. The addition of boron ions to the system was direct. The desired weight percent of the $P_2O_5$ in the final (ceremized) material is presented in Table 1. The addition of the commodifying ions corresponded to the molar P:Al or P:B ratio equal to 1:1 for the lowest phosphate concentration.

The obtained materials underwent the low-temperature treatment up to 150 °C, during which the evaporation of the solvent went along with the completion of the polycondensation process. Afterwards, the obtained bulk polymeric samples underwent high-temperature treatment at 800 °C in the protective atmosphere on noble gas (argon) for 30 minutes (heating rate 5 °C/min, free cooling).

*Tab. 1. Concentrations of the modifiers - desired weight percent (wt.%) of the corresponding oxide ($P_2O_5$) in the final material*

| Series | (2-diethylphosphatoethyl)-triethoxysilane | | | triethyl phosphite | | | Co-modifier |
|---|---|---|---|---|---|---|---|
| | **1** | **2** | **3** | **1** | **2** | **3** | |
| **A** | 2.5% | 5% | 7.5% | — | — | — | — |
| **B** | — | — | — | 2.5% | 5% | 7.5% | |
| **C** | 2.5% | 5% | 7.5% | — | — | — | aluminium tri-sec-butoxide |
| **D** | — | — | — | 2.5% | 5% | 7.5% | |
| **E** | 2.5% | 5% | 7.5% | — | — | — | trimethyl borate |
| **F** | — | — | — | 2.5% | 5% | 7.5% | |

Thermal analysis (thermogravimetry (TGA) and differential scanning calorimetry (DSC)) was performed on dry polymeric samples after low-temperature treatment with NETZSCH STA 449 F3 analyzer. A 5 °C/min heating rate in a protective argon atmosphere up to 1100 °C was used.

Extensive structural studies were initiated with the XRD analysis with X'Pert Pro Philips (PANalytical) using CuK$_{\alpha 1}$ radiation source in Bragg-Brentano geometry with 0.008 deg step size in the 5−90 deg 2θ range. Polymeric (xerogel) and ceramic powder samples were analysed with the Debye-Scherrer-Hull (DSH) method. Moreover, based on the position of the double halo on the polymers diffractograms, it was possible to calculate the average size of the synthesised silsesquioxane ladders and the distance between them, according to Equation 1 [21]:

$$d_{1\ or\ 2} = \frac{\lambda}{2\sin\theta} \qquad \text{(Eq. 1)}$$

Where:     $\lambda\ [\text{Å}]$ – X-ray lamp length (for CuK$_{\alpha 1}$=1,5406 Å)
            $2\theta\ [deg]$ - reflection angle
            $d_1$ – the distance between the ladders; $d_2$ – width of the ladder

Fourier-transformed infrared spectroscopy (FTIR) measurements were performed with VERTEX 70v (Bruker) apparatus in the middle infrared (MIR) range (400-4000 cm$^{-1}$) on both polymers and ceramic samples. Transmission mode with the use of KBr pellets was used. Spectra were acquired with 128 scans and 4 cm$^{-1}$ resolution. Subsequently, spectra were subjected to basic post-processing including baseline correction (using polynomial function) conducted in OPUS 7.2 software that was also used to record the spectra. Due to the amorphous character of the samples, hence high width of the bands, the post-treatment decomposition of the spectra was required, especially of the ceramic samples. Also, the sections of the polymeric samples spectra were decomposed for a more detailed analysis of the region of interest covering the bands characteristic for vibrations in ladder-like silsesquioxanes. The decomposition process was based on the Voight function and the Levenberg-Marquardt algorithm based on the least squares method. The bands' positions were revealed using the second derivative of the spectra.

Raman microspectroscopic analysis was performed with WITec Alpha M+ spectrometer (WITec, Germany) equipped with a 488 nm diode laser and 50x long-focal lens. Spectra were collected using WITec Control FIVE software, based on 2 scans with 20 s accumulation time. In the case of some of the ceramic samples modified with phosphate and boron ions, it was impossible to collect data due to the multiple factors described in detail in the following part of the article.

Magic Ange Spinning Nuclear Magnetic Resonance (MAS-NMR) spectra were recorded on the Apollo (Tecmag) spectrometer console equipped with the magnetic field of 7 T produced by a 300 MHz superconducting magnet (Magnex) with a gap at the level of 89 mm. The Bruker HP-WB MAS probe with a 4 mm zirconia rotor was used.

### 3. Results and discussion
### 3.1. Thermal analysis of polymeric precursors

TGA analysis enabled to indicate the evolution of the mass of the samples and crucial temperature point when the mass evolution is finished. DSC depicted both exo- and endothermal transformations, most interestingly not connected with the mass change, such as crystallisation, since one wants to avoid it. The obtained curves are typical for the preceramic precursor for silicon oxycarbide-based materials [22]. The difference between them and the one obtained before (with no dopants) is subtle [22].

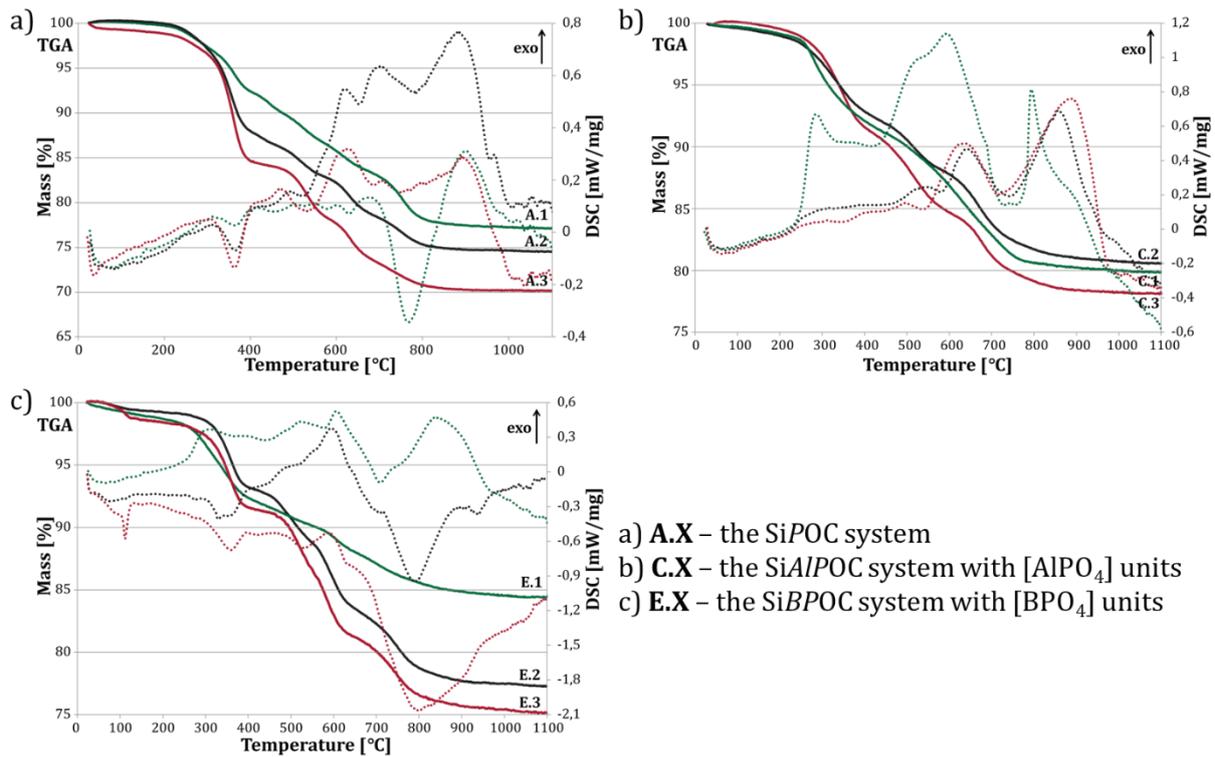

a) **A.X** – the SiPOC system
b) **C.X** – the SiAlPOC system with [AlPO₄] units
c) **E.X** – the SiBPOC system with [BPO₄] units

*Figure 1. TGA (straight lines) and DSC (dotted lines) curves of the polymeric precursors of chosen series: (a) A: SiPOC system, (b) C - SiAlPOC system and (c) E – the SiBPOC system*

The thermal evolution of the polymers into the ceramics (Fig. 1) might be described in a few stages depending on the main occurring processes. The first one is connected with water and solvent residue evaporation and takes place below 250 °C. The mass loss in this area is negligibly low (approx. 2%), because of the polymeric samples pre-treatment. It is more connected with the side products of the ongoing polymerisation process (which will be finalised at the ceramization point) [23, 24]. Between 250 °C and 800 °C, the primary polymer-to-ceramic transformation occurs. The significant mass loss is bound with a broad exothermic effect. The ongoing processes include decomposition of organic groups, evaporation of the unreacted monomers and lightweight oligomers [25], and redistribution of Si-O and Si-C bonds [26, 27]. Here, the matrix of the silicon oxycarbide is created along with the free carbon phase. Between 750-790 °C endothermic pick indicates the nucleation process for all samples. Above 800 °C processes connected with phase separation and crystallisation are visible as an exothermic effect. It was not observed only in the case of SiBPOC samples of higher phosphate concentration (Fig. 1c).

The general mass loss is between 14.7% for SiBPOC system with the lowest P content and 32.1% for SiPOC system with the highest P content. The great majority of the loss occurs below 800 °C. Above that temperature, the loss is negligible (below 1.5%). That indicates the most favourable ceramization temperature. No significant influence of the dopants was observed. Thus, there is no general correlation between the presence of phosphate, aluminum and boron ions and thermal effects besides the lack of crystallisation peak for SiBPOC E.2 and E.3 samples (Fig. 1c).

### 3.2. X-ray diffraction of polymeric precursors and ceramized powders

Obtained diffractograms of the polymeric powder samples (Fig. 2a, b, c) show a double halo profile, characteristic of the ladder-like structures. Based on the positions, it was estimated that the width of the ladders varies between 4.05 and 4.22 Å, where the distance between them lies between 8.08 and 8.72 Å [21]. These agree with the results obtained for the unmodified silicon oxycarbide, yet the ladders were slightly broadened (less than 5% compared with the unmodified polymer). In the case of the samples containing boron ions, additional peaks are visible due to the boron oxide (2θ 14.6° and approx. 28°) [28, 29]. The remaining minor signals (42, 49, 72 and 87°) originated in the brass substrate used during measurements.

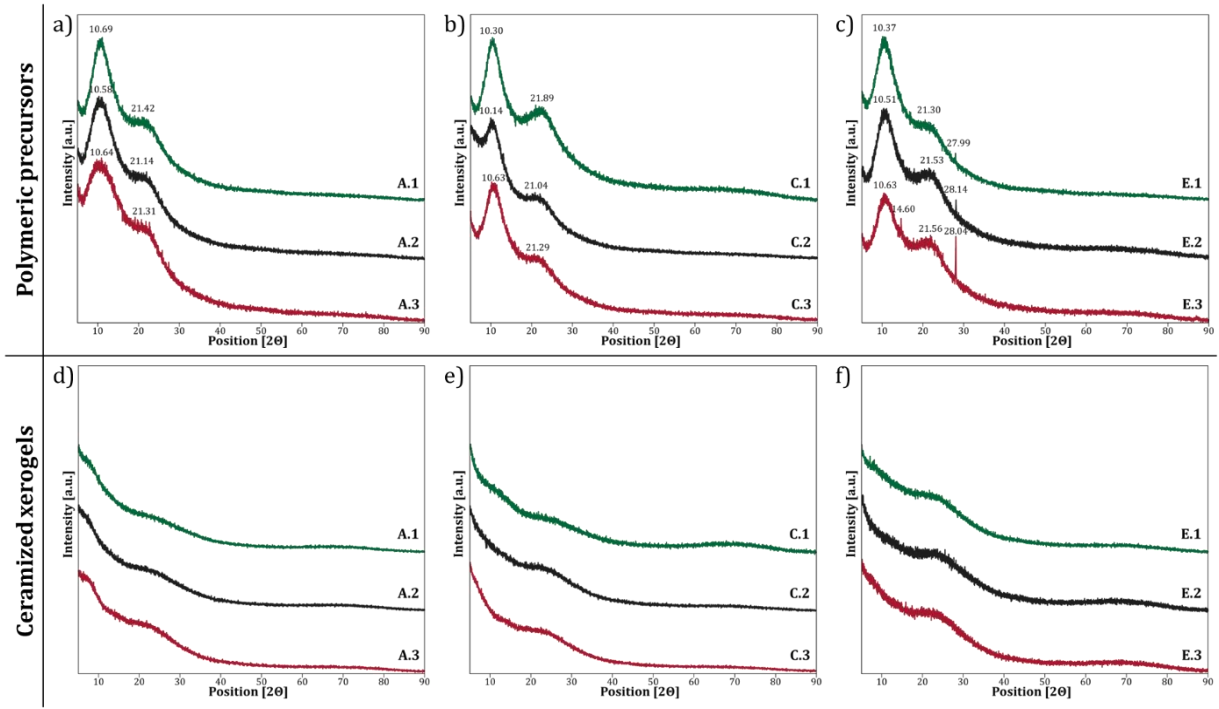

*Figure 2.XRD patterns of obtained materials before (a, b, c) and after (d, e, f) ceramization*

After the ceramization process, only a single halo is visible (Fig. 2d, e, f). That indicates the complete decomposition of the polymeric ladder-like structure and boron oxide.

### 3.3. *Fourier-transformed infrared spectroscopy of polymeric precursors*

The main task of the FTIR was to answer the question of how the cationic modification influenced the structure of the preceramic polymers (Figs. 3-5). The first point was to confirm the presence of the Si-C bonds. It was done by identification of the bands in the region between 700-900 cm$^{-1}$: stretching vibrations of Si-C bonds connected with bending vibrations of C-H [1, 22, 19, 30]. However, the analysis of the obtained spectra is complex due to the overlapping phenomenon of P-O, B-O and Al-O bands with the Si-O bands in the SiOC matrix. The first of such overlapping regions is at approx. 560 cm$^{-1}$, where bands of the bending modes of P-O, B-O bonds, along with Si-O-Al stretching vibrations [19, 17, 31] overlap with vibrations of four-member Si-O rings [1, 22, 19, 30, 32]. The intensity of the bands is moderated with the varying quantity of the added phosphate ions, indicating a differing amount of the P-O, B-O, and Si-O-Al bonds and four-membered Si-O rings.

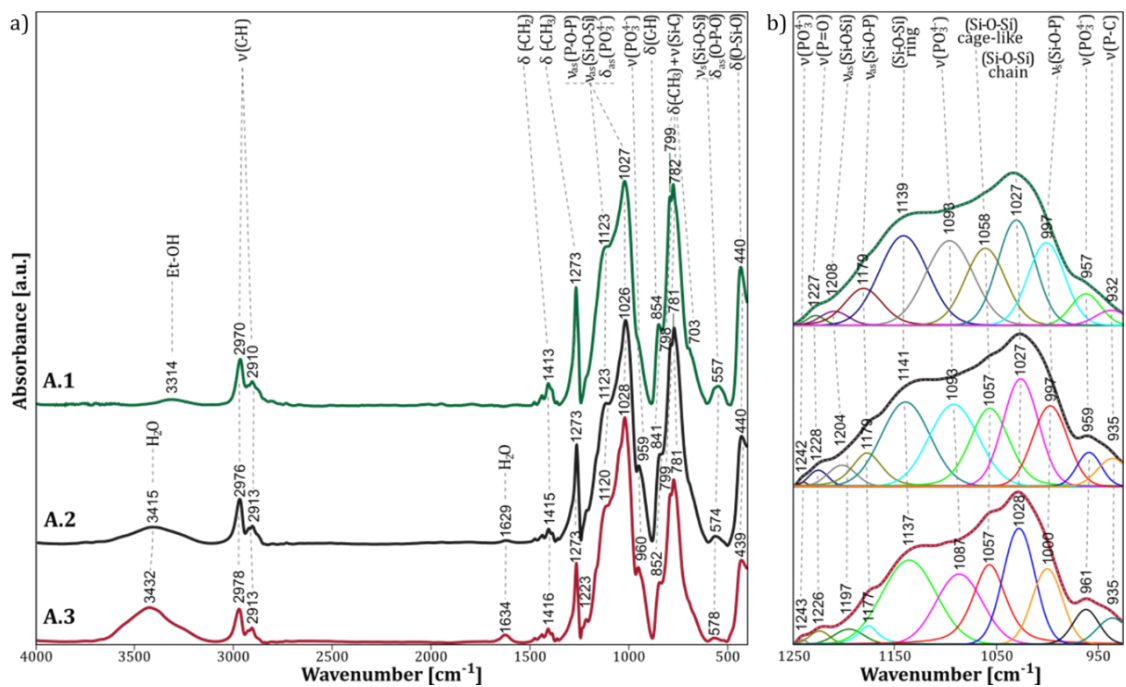

*Figure 3. Infrared spectra of the polymeric precursors – ladder-like silsesquioxanes modified with phosphate ions (based on (2-diethylphosphatoethyl)-triethoxysilane): full-range spectra (a) followed by decomposition of the chosen region to component bands (b)*

The key for identifying the structure of ladder-like silsesquioxanes is the spectral region between 1000 and 1200 cm⁻¹ with the dominant bands at approx. 1030 and 1100 cm⁻¹ responsible for stretching vibrations of Si-O-Si bridges, more precisely network and ring-like vibrations of T structural units ([SiO₃C]) in the ladder, respectively [1, 22, 19, 33, 34, 35, 36]. It was necessary to perform decomposition of the spectra to the sub-bands, because in this region, the stretching vibrations of the Si-O-P and P-O-P bonds are also present and partially overlap [15].

As mentioned above, the range between 900 and 1250 cm⁻¹ comprises multiple partially overlapping bands. Its analysis requires component band decomposition. This operation is necessary to observe critical stretching vibrations of Si-O bonds in ladder-like silsesquioxanes in numerous configurations: chain-like (1030 cm⁻¹), ring-like (1130 cm⁻¹) and cage-like (1060 cm⁻¹). Last to be mentioned is the asymmetric stretching vibration of silicon-oxide bridges connecting ladders' chains [37]. The crucial P-O bands of [PO₄] units and Si-O-P bridges at 960 and 1170 cm⁻¹, respectively, might also be found [15]. In the A and C series spectra, additional two profound bands might be found at 935 and 1242 cm⁻¹. They are connected with the use of the organophosphate substrate and are responsible for P-C and P=O band vibrations [38, 15, 39]. Moreover, their intensities are dependent on the amount of introduced precursor and rise from X.1 to X.3, where X=A, C, E. Besides using the same precursor in the series E, the bands were not observed or barely visible. In the case of series C additional band at 1050 cm⁻¹ was observed. It is linked with multiple oxygen bridges, including P-O-P and Si-O-Si [40], but because it is unique for the samples containing aluminium, it may be recognised as the Si-O-Al band [41]. For series A and C, the presence of $PO_4^{3-}$ was recognised by the presence of the band at approx. 960 cm⁻¹ that is responsible for the stretching vibration of the unit [15].

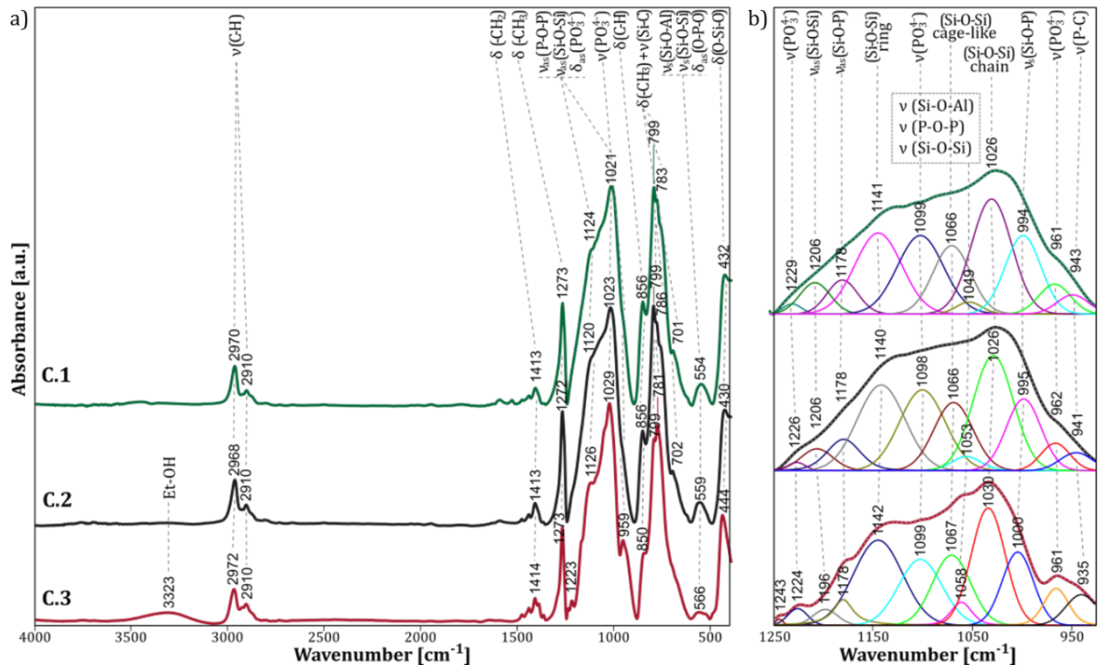

*Figure 4. Infrared spectra of the polymeric precursors – ladder-like silsesquioxanes modified with phosphate ions (based on (2-diethylphosphatoethyl)-triethoxysilane) and co-dopped with aluminium ions: full-range spectra (a) followed by decomposition of the chosen region to component bands (b)*

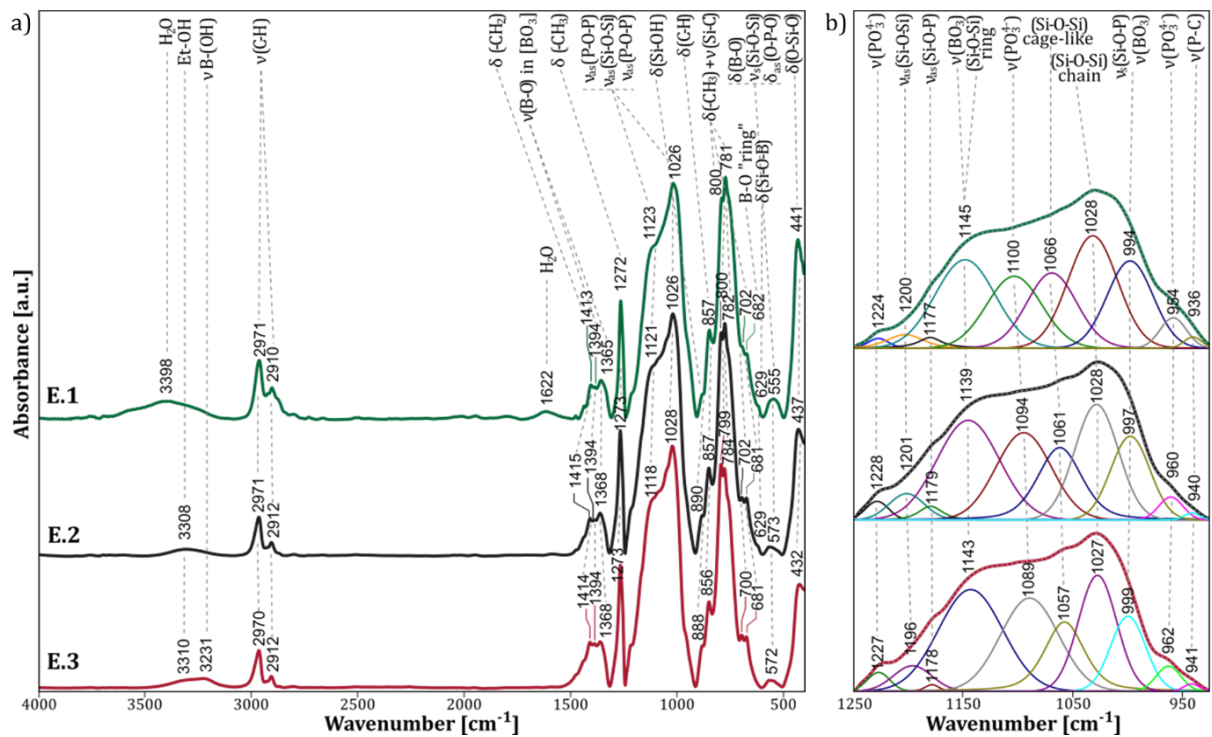

*Figure 5. Infrared spectra of the polymeric precursors – ladder-like silsesquioxanes modified with phosphate ions (based on (2-diethylphosphatoethyl)-triethoxysilane) and co-dopped with boron ions: full-range spectra (a) followed by decomposition of the chosen region to component bands (b)*

In the case of the polymeric materials modified with boron (Fig. 5) multiple additional bands occurred: Si-O-B bending at 630 cm⁻¹ [40] and ring B-O vibration at 682 cm⁻¹ [38]. In the 1360-1400 cm⁻¹ range, stretching vibrations of the B-O bonds in the [BO₃] units are present [15, 40, 38]. It includes the band at 1394 cm$^{-1}$ responsible for the isolated borate ions $BO_3^{3-}$ [40]. In some samples, the band at 3220 cm$^{-1}$ of the stretching vibration of the hydroxyl group in B-OH units is also present [40]. The presence of the Si-OH group [19, 42] indicates that the polycondensation process is not fully finished (similarly to B-OH).

*Tab. 2. Values of the integral intensities ratio of the characteristic ring and chain bands correlated with the chain-like and cage character of the obtained silsesquioxanes*

| Series | $R = \dfrac{I_{1130}}{I_{1030}}$ | Series | $R = \dfrac{I_{1130}}{I_{1030}}$ | Series | $R = \dfrac{I_{1130}}{I_{1030}}$ |
|---|---|---|---|---|---|
| **A.1** | 1.03 | **C.1** | 0.82 | **E.1** | 0.94 |
| **A.2** | 1.08 | **C.2** | 0.88 | **E.2** | 1.24 |
| **A.3** | 1.24 | **C.3** | 1.05 | **E.3** | 1.48 |

As the method of materials' characterisation, the ring (at 1130 cm$^{-1}$) to chain (at 1030 cm$^{-1}$) bands' integral intensities ratio was calculated (Table 2). The obtained values might show the character of the obtained compound. And here, if the received value is above 1, dominantly cage-like structures connected by singular ladders was received. With a ratio strongly below 1, the network is chain-like with a limited number of cages. With a ratio close to 1, the structure is intermediate, and the material consists mainly of ladder-like structures with singular cage-like structures. Since the results vary between 0.8 and 1.5, the mixed cage-chain structures were obtained with the character toward chain-like structures for the C series and cage-like for E. However, it is important to recognize that in the case of boron-containing series E, the indicated bands overlap with $BO_3$ unit vibrations [40]. Nevertheless, it might be noticed that inside the series ratio's value increases with increasing phosphate ions concentration.

### 3.4. Raman spectroscopy of polymeric precursors

As the complementary method to FTIR, Raman spectroscopy was used. In the Raman spectra (Figs. 6-8), the characteristic bands for ladder-like silsesquioxanes might be found with the crucial bands connected with Si-C bond vibrations at 180, 700 and 780 cm$^{-1}$ [19]. Bands corresponding to bending vibrations of O-P-O bridges are observed at 320 and 590 cm$^{-1}$ [40]. Stretching vibrations of P-O are at 950 and 1010 cm$^{-1}$ [31]. Again some bands strictly connected with the phosphate precursor (i.e. P=O stretching at 1160 and 1230 cm$^{-1}$) can be followed [43, 44]. The presence of aluminium ions was confirmed through bands at 700 and 950 cm$^{-1}$ related to vibrations of [$AlO_4$] and Al-(O-H) units, respectively [19, 45]. The profiles of spectra of xerogels from Si*P*OC and Si*AlP*OC systems are relatively similar. With the addition of boron, the spectrum shape changes significantly. Bands of the partially or entirely hydrolysed boron monomer might be observed at 880 cm$^{-1}$ ($B(OH)_3$), 795 cm$^{-1}$ ($B(OH)_2(OR)$) and in the range 720-760 cm$^{-1}$ ($B(OH)(OR)_2$ and $B(OR)_3$) [46]. The polycondensation of the boron ions was revealed by the band's presence at 1020 cm$^{-1}$ connected with stretching vibrations of B-O-B and B-O-Si bridges [40].

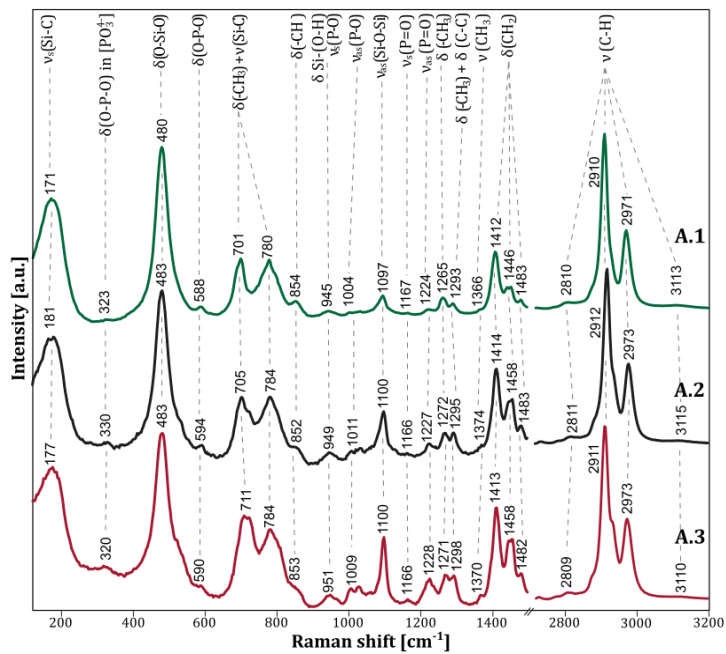

*Figure 6. Raman spectra of xerogels from the SiPOC series*

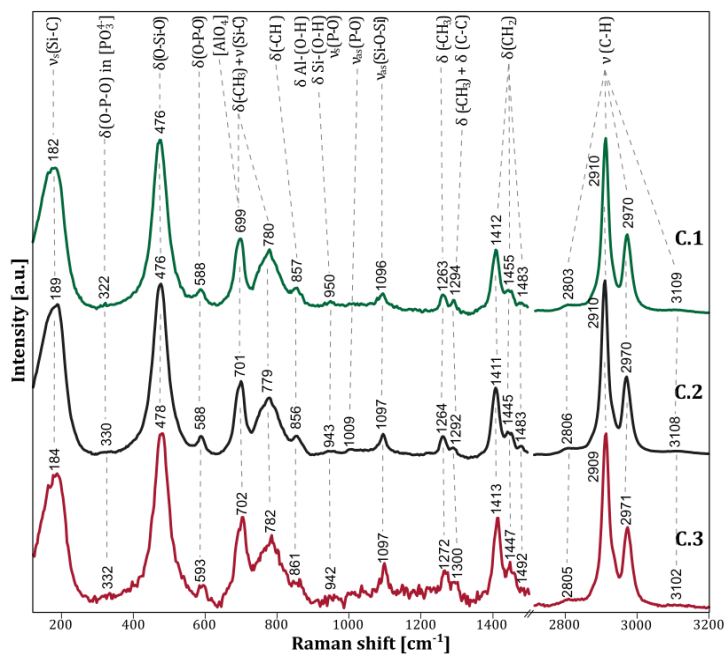

*Figure 7. Raman spectra of xerogels from the SiAlPOC series*

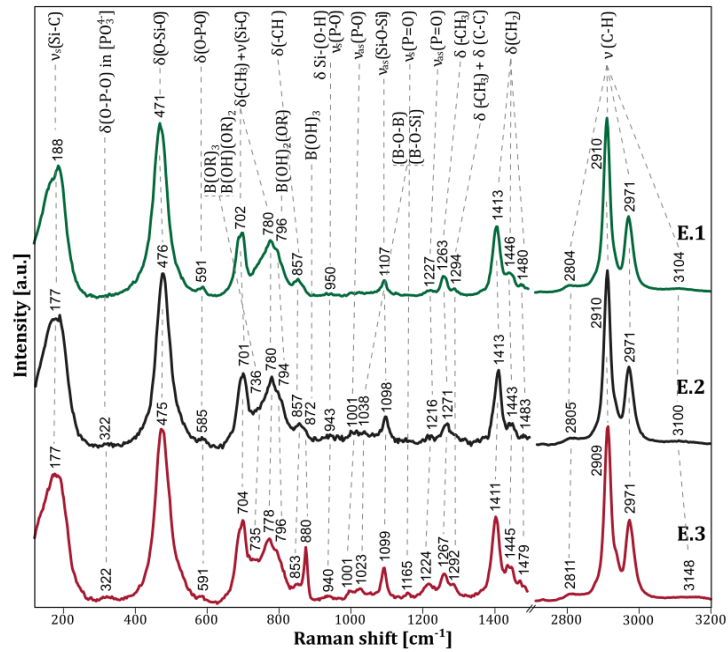

*Figure 8. Raman spectra of xerogels from the SiBPOC series*

### 3.5. *Fourier-transformed infrared spectroscopy of ceramic powders*

With FTIR, it was possible to confirm the effective preparation of modified silicon oxycarbide amorphous materials. The spectra of SiOC-based materials (Fig. 9-11) are similar to one of amorphous silica [47]. It is characterised by overlapping broad bands of high width at half maximum. Therefore, to analyse the spectra, it is essential to decompose them. On such prepared spectra, it is possible to find a band at 840 cm$^{-1}$ of the stretching vibrations of the Si-C bond. Among numerous bands analogical to amorphous silica the following findings should be highlighted: Si-O bridges (< 500 cm$^{-1}$), 3-, 4- and 6-membered rings (500-700 cm$^{-1}$) and stretching Si-O 750-1200 cm$^{-1}$, Si=O 1200 cm$^{-1}$, Si-O-Si (1040 and 1120 cm$^{-1}$) and broken Si-O$^-$ bridges 980 cm$^{-1}$ [47, 48]. The bands at 1365 and 1278 cm$^{-1}$ show that the ceramization process is not completed [19]. The bands at 1105 cm$^{-1}$ and 600 cm$^{-1}$ might confirm the presence of the modifiers [31, 40, 49]. In the case of Si*Al*POC materials (Fig. 10), the band of asymmetric stretching vibrations of Si-O-Al at 1010 cm$^{-1}$ proves that Al ions built into the network. Aluminium presence is also proven by the bands at 440, 580 and 908 cm$^{-1}$, responsible for O-Al-O bending, octahedral alumina and Al-OH stretching, respectively [19].

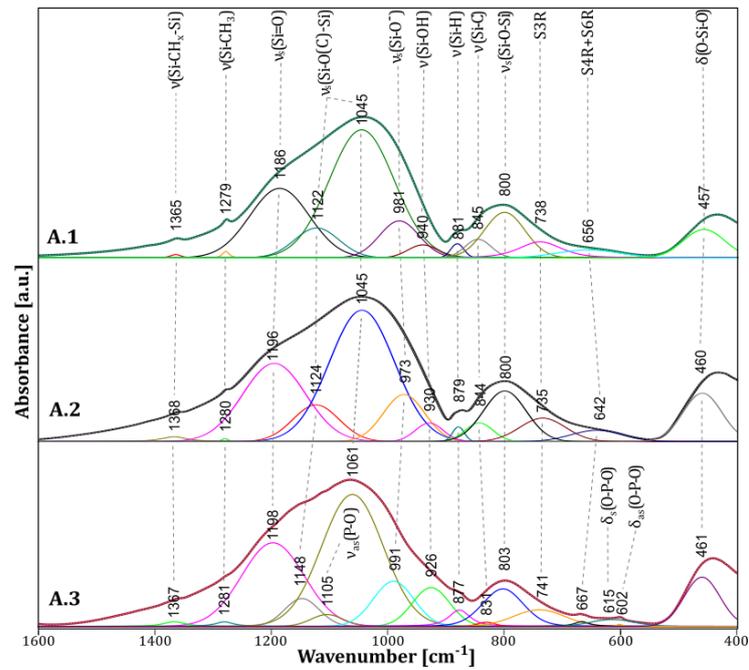

*Figure 9. Decomposed FTIR spectra of ceremized SiOC-based modified amorphous materials modified with phosphate ions*

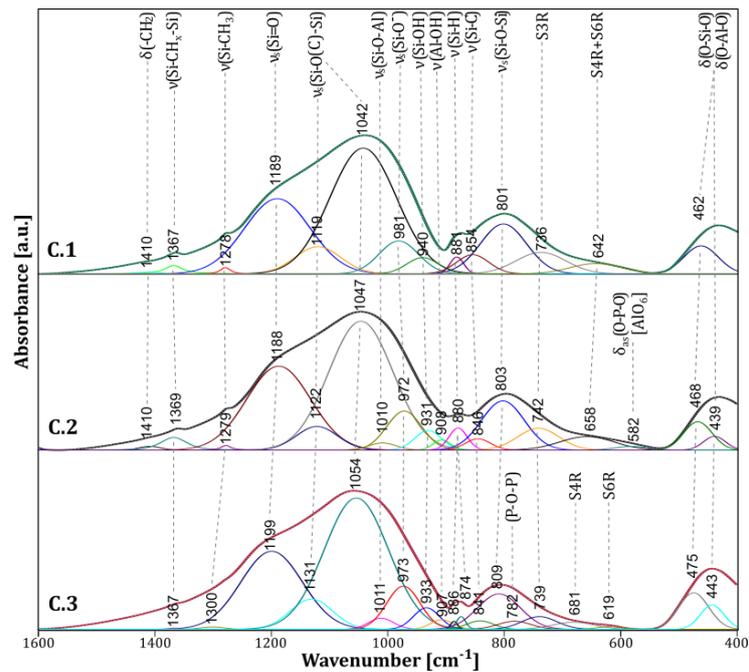

*Figure 10. Decomposed FTIR spectra of ceremized SiOC-based modified amorphous materials modified with phosphate and co-doped with aluminium ions*

Analysis of materials from Si*BP*OC is more complex (Fig. 11). Here, the energy of the B-O bond is similar to one of Si-O and P-O. Notwithstanding, it was possible to identify certain bands at 1100 and 627 cm$^{-1}$ suggesting the incorporation of boron ions as the network-creating element [40].

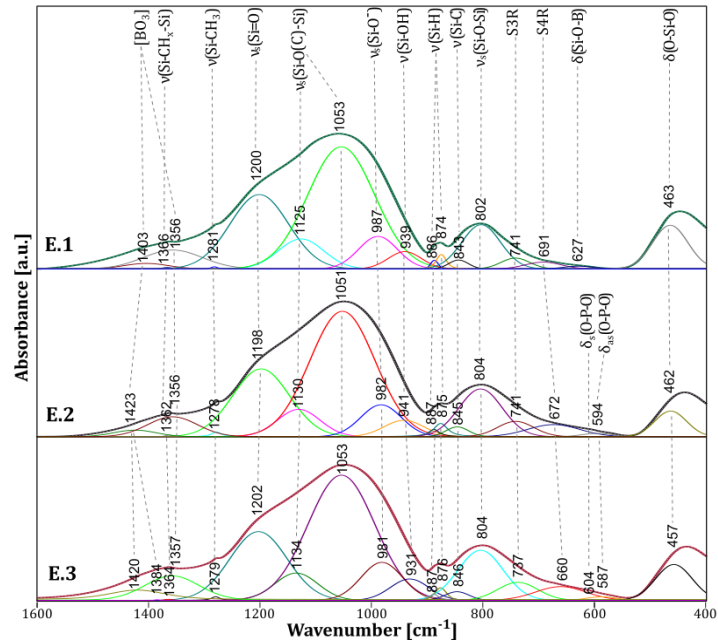

*Figure 11. Decomposed FTIR spectra of ceremized SiOC-based modified amorphous materials modified with phosphate and co-doped with boron ions*

### 3.6. *Raman spectroscopy of ceramic powders*

Raman analysis of SiOC-based materials is difficult due to the fluorescence phenomenon. The analogical effect was present in the modified materials, most pronounced for the Si*BP*OC system (Fig. 12). Obtained spectra are composed of two strong bonds at approximately 1350 cm$^{-1}$ and 1600 cm$^{-1}$, both originating from the turbostratic carbon phase [1]. The first mentioned, the so-called D band, corresponds to symmetric breathing vibrations of hexagonal carbon rings and is associated with defects. The second band, the so-called G, is connected with the stretching C-C vibrations in the graphene sheets of atoms in sp$^2$ hybridisation [27, 50, 1, 51, 52]. The ratio between integral intensities of D and G bands (designated based on fitting) was calculated to indicate the level of structural disorder in the carbon phase. The obtained $I_D/I_G$ values are presented on the Fig. 12c. In case of the A series (Si*P*OC) with higher phosphate ions concentration, the trend of rising ratio, hence increasing disorder, is visible. However, for the C series, the ratio is relatively constant. The effect of phosphate ions concentration in the series A is visible in the shift of the position of the G band towards lower values, indicating amorphisation of the carbon phase [51, 53]. Registering the spectra of E.2 and E.3 samples was impossible. It might be connected with the influence of the boron ions on the character of the carbon phase in silicon oxycarbide-based materials [18].

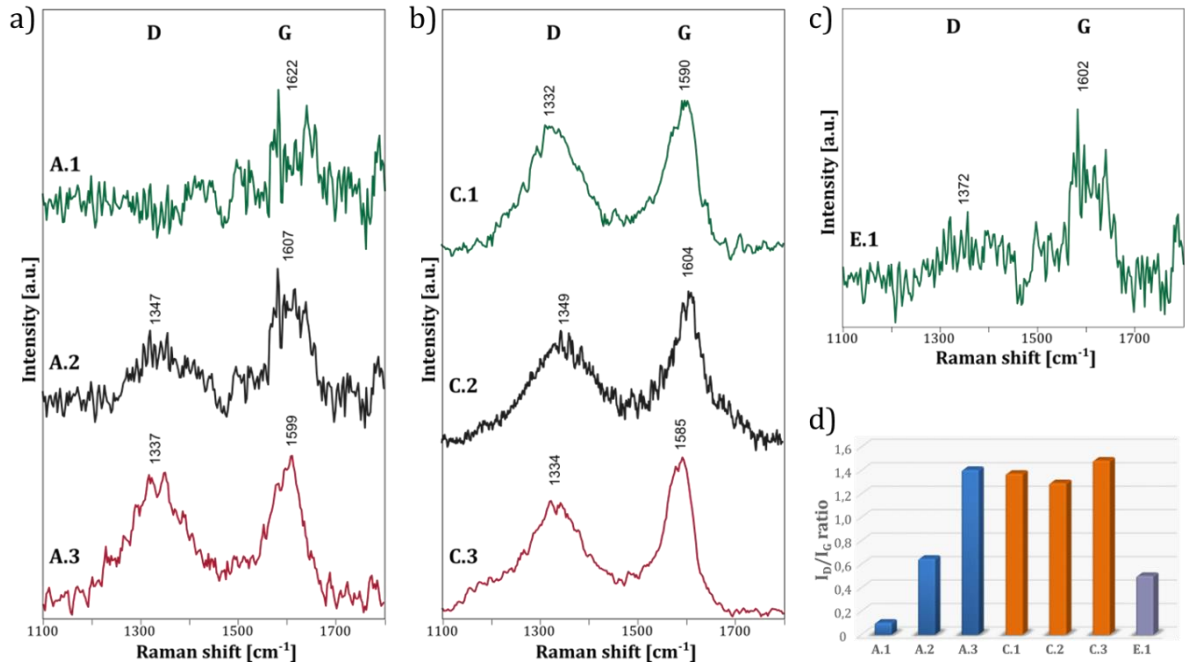

*Figure 12. Raman spectra of the modified SiOC-based materials with visible D and G bands characteristic for defected graphite: (a) SiPOC system, (b) SiAlPOC system, (c) SiBPOC system and (d) calculated $I_D/I_G$*

### 3.7. *Magic angle spinning nuclear magnetic resonance of SiAlPOC sample*

Selected ceramic powder sample from the Si*Al*POC system (C.1) underwent MAS-NMR measurements on $^{29}$Si, $^{27}$Al and $^{31}$P nuclei (Fig. 13). Silicon analysis (Fig. 13a) showed three types of the Si-O structural units. T ([SiO$_3$C]) and D ([SiO$_2$C$_2$]) units were directly transferred from the preceramic precursors (ladder-like polysilsesquioxanes), while Q structural units ([SiO$_4$]) occurred as the result of bonds redistribution during the ceramization process [25]. The profile of the obtained signals is characteristic for amorphous silicon oxycarbide [27, 19, 54]. In the case of the $^{27}$Al nucleus (Fig. 13b), it was possible to decompose a broad spectrum to the component bands demonstrating the presence of aluminium in different coordination. The most pronounced [AlO$_6$] octahedron suggests that part of alumina acted as a glass matrix modifier, creating broken silicon-oxygen bridges. The network-forming tetrahedral coordinated alumina ([AlO$_4$]) was detected in smaller quantities. Moreover, the intermediate form of alumina ([AlO$_5$]), present on the interface between domains, was found [19, 55]. The spectra of $^{31}$P had a very low signal-to-noise ratio making the analysis more challenging. The observed signal close to 3 ppm was in between the expected signals from Q$^0$ and Q$^1$ units corresponding to the PO$_4^{3-}$ and P$_2$O$_7^{4-}$, respectively [56, 57], suggesting an intermediate character of the nuclei.

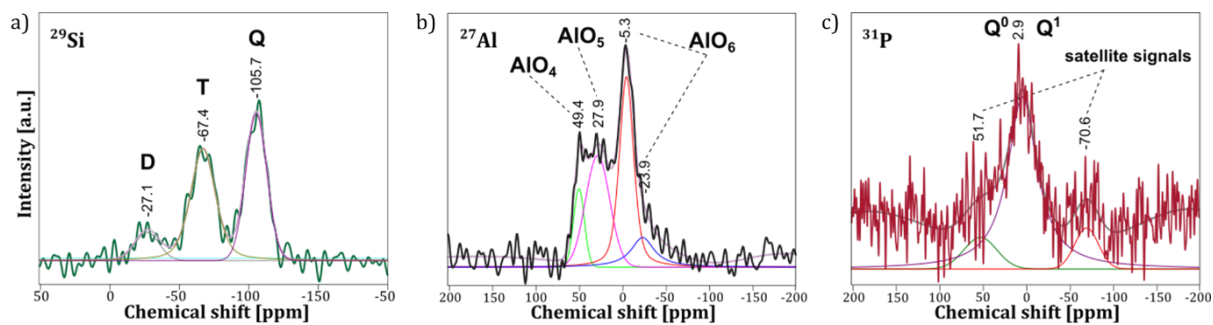

*Figure 13. MAS-NMR spectra of the ceramic C.1 sample from SiAlPOC system*

**Conclusions**

The presented study showed an effective methodology for preparing silicon oxycarbide-based amorphous materials from Si*P*OC, Si*AlP*OC and Si*BP*OC systems. Structural study with XRD, FTIR and Raman spectroscopy showed a ladder-like character of the preceramic precursors with built-in modifying phosphate, aluminium and boron ions. The crucial bonds were preserved during thermal treatment and depicted with the spectroscopic method. Moreover, the boron oxide in the E series was effectively decomposed. It was essential to decompose FTIR spectra to the component bands due to amorphous character of received glasses, the similarity of the matrix to the amorphous silica as well as the close positions of bands corresponding to the modifiers bonded with oxygen. It was also shown that the impact of the modifiers on the glass matrix is noticed by the subtle changes of spectra profiles with new bands' presence. The exception is the drastic change in the character of the turbostratic graphite-like phase in the Si*BP*OC. It is expected (and will be further explored) for the presented modifications to significantly impact the properties of the obtained coatings.

*Acknowledgement*

The work was financed by the NCN project No 2019/35/B/ST5/00338. Research project supported by the program "Excellence initiative – research university" for the AGH University of Science and Technology. Magdalena Gawęda acknowledges support by the NOMATEN MAB grant for partly covering the personnel costs, agreement no. MAB PLUS/2018/8.

The authors want to acknowledge the head of the X-ray diffraction laboratory prof. Bartosz Handke for conducting the measurements.